\documentclass[fleqn,10pt]{wlscirep}
\usepackage[utf8]{inputenc}
\usepackage[T1]{fontenc}
\usepackage{hyperref}
\usepackage{amsmath,amssymb,amsthm,bbm, mathtools} %
\usepackage{mathrsfs}
\usepackage{xcolor}
\usepackage{multirow}
\usepackage{cite}
\usepackage{pifont}
\usepackage{graphicx}
\usepackage[normalem]{ulem}

\usepackage{color}      % use if color is used in text
\usepackage{ulem}

\title{Protected valley splitting against interface disorder toward scalable silicon electron spin qubits}

\author[1,+]{Yang Liu}
\author[2,+]{Gang Wang}
\author[1]{Shan Guan}
\author[1,*]{Jun-Wei Luo}
\author[1]{Shu-Shen Li}
\affil[1]{State Key Laboratory of Semiconductor Physics and Chip Technologies, Institute of Semiconductors, Chinese Academy of Sciences, Beijing 100083, China}
\affil[2]{School of Physics, Ningxia University, Yinchuan 750021, China}

\affil[*]{jwluo@semi.ac.cn}

\affil[+]{these authors contributed equally to this work}

\begin{abstract}
 Regardless of various material design strategies, experimentally achieving substantial and controllable valley splitting in Si/SiGe quantum wells remains a central challenge for ensuring high gate uniformity. This difficulty arises from unavoidable atomic-scale disorder at the interface, caused by alloy randomness, which suppresses valley splitting and, more critically, induces large variations. Here, we demonstrate that CMOS-compatible uniaxial strain can substantially enhance valley splitting, rendering it immune to interface disorder. Atomistic pseudopotential calculations show that uniaxial strain linearly restores the valley splitting suppressed by interfacial disorder, with a large enhancement rate, while keeping disorder-induced variations within a narrow distribution. We reveal that uniaxial strain introduces a new coupling channel between bulk valleys in adjacent Brillouin zones through a small momentum transfer, which markedly reduces the susceptibility of valley splitting to interfacial disorder. These findings establish a viable route to improve gate uniformity in silicon-based spin qubits, paving the way for scalable quantum processors.
\end{abstract}
\begin{document}

\flushbottom
\maketitle
%\thispagestyle{empty}

\section*{Introduction}
 
Electron spins confined in gate-defined quantum dots in Si/SiGe quantum wells (QWs) present a promising platform for developing large-scale fault-tolerant quantum computers~\cite{PhysRevA.57.120, RevModPhys.79.1217, RevModPhys.85.961, zhang2019semiconductor, RevModPhys.95.025003}, facilitated by the zero-nuclear-spin isotopes that enable spin lifetimes as long as 10 s~\cite{tyryshkin2012electron} and by mature microelectronics technology that permits ultra-large-scale integration. The recent demonstration of two-qubit gate fidelity exceeding the 99\% fault-tolerance threshold for quantum computation~\cite{xue2022quantum, noiri2022fast, mills2022two, pengDiverseMethodsPractical2026, zhouExchangeInteractionNeighboring2024} has shifted the focus toward improving gate uniformity for scalable quantum processors. Many of the essential qubit metrics are closely linked to the material quality of the underlying Si/SiGe QWs, with valley splitting being a particularly vital parameter~\cite{scappucciCrystallineMaterialsQuantum2021}, since it measures the energy separation between two lowest (originally degenerate) valley states and therefore must be sufficiently large to enable high-fidelity spin qubit initialization, readout, control, and shuttling~\cite{yang2013spin, PhysRevB.89.075302, PhysRevB.90.235315, PhysRevLett.121.076801, PhysRevB.83.165322, PhysRevB.86.035302, zhao2019coherentelectrontransportsilicon, PRXQuantum.2.040358, tariq2022impact}. However, recent experimental studies have revealed that the progress of Si electron spin qubits toward scalability is largely hindered by both the smallness and variability of valley splitting~\cite{paqueletwuetzAtomicFluctuationsLifting2022,liu2022zoo,liu2024progress}, which manifest in most qubit devices as insufficient splitting for uniform and reliable operation~\cite{philips2022universal,weinstein2023universal,degli2024low}.

The magnitude of the valley splitting ($E_\mathrm{VS}$) in Si/SiGe QWs is highly sensitive to the atomistic structure of the interface. In particular, alloy disorder or interface steps are well documented to substantially reduce the valley splitting from ideal values to typically in the range of 10-100 $\mu$eV~\cite{10.1063/1.3569717, 10.1063/1.3666232, 10.1063/1.4922249, PhysRevApplied.13.034068, PhysRevB.95.165429, PhysRevB.98.161404, PhysRevLett.119.176803, goswami2007controllable, PhysRevLett.125.186801, PhysRevApplied.11.044063}. Early atomistic simulations~\cite{zhang2013genetic} further revealed that the atomic fluctuations in the alloy barrier introduce substantial variations in this already reduced valley splitting, a finding recently confirmed by experimental observations showing a wide spread across quantum dot devices~\cite{paqueletwuetzAtomicFluctuationsLifting2022}. Several atomic-scale strategies have been explored to boost valley splitting in Si/SiGe QWs to a few meV, including superlattice barriers~\cite{zhang2013genetic, PhysRevB.105.165308}, narrow Ge spikes~\cite{PhysRevB.104.085406}, and wiggle wells~\cite{PhysRevB.106.085304, mcjunkin2022sige, gradwohlEnhancedNanoscaleGe2025}. However, no matter which current material design strategy is adopted, achieving substantial and controllable valley splitting in Si/SiGe QWs remains a major challenge in practice. This is because unavoidable interface disorder in the alloy barrier, together with the limited growth precision imposed by atomic interdiffusion, severely undermines their effectiveness, continuously suppressing the attainable valley splitting and introducing significant fluctuations~\cite{paqueletwuetzAtomicFluctuationsLifting2022,PhysRevB.108.125405}. This destructive sensitivity to disorder has so far prevented these strategies from delivering meaningful improvements, posing a major obstacle to scalable Si spin qubits. Consequently, a strategy that can suppress the effects of interface disorder while deterministically enhancing the valley splitting is urgently needed.

In this work, we demonstrate through atomistic semi-empirical pseudopotential (SEPM) calculations that uniaxial strain can substantially enhance the immunity of valley splitting to interfacial disorder. A microscopic theoretical model is developed by incorporating shear strain into the conventional valley splitting framework to elucidate the underlying mechanism. We show that shear strain activates a new coupling channel between bulk valleys in adjacent Brillouin zones involving a small momentum transfer, thereby markedly reducing the sensitivity of valley splitting to atomic-scale disorder. Notably, the concept of enhancing valley splitting by shear strain in QWs with ideal interfaces dates back to the 1970s~\cite{ohkawa1977theory} and has recently regained interest both theoretically and experimentally~\cite{SVERDLOV20081861, sverdlov2011strain, PhysRevLett.133.037001, woods2024coupling, thayilTheoryValleySplitting2025, noborisakaValleySplittingExtended2024, marcoglieseFabricationCharacterizationMechanical2025}. However, whether this enhancement survives in the presence of unavoidable interface disorder has remained unclear. Our findings provide not only fundamental insights into shear-strain modulation of valley splitting but also a practical strategy to mitigate disorder-induced variability in Si-based devices, paving the way for scalable and robust quantum computing architectures.

\section*{Results}

\subsection*{Atomistic simulation results}
\begin{figure}[t]
	\centering
	\includegraphics[width=0.9\textwidth]{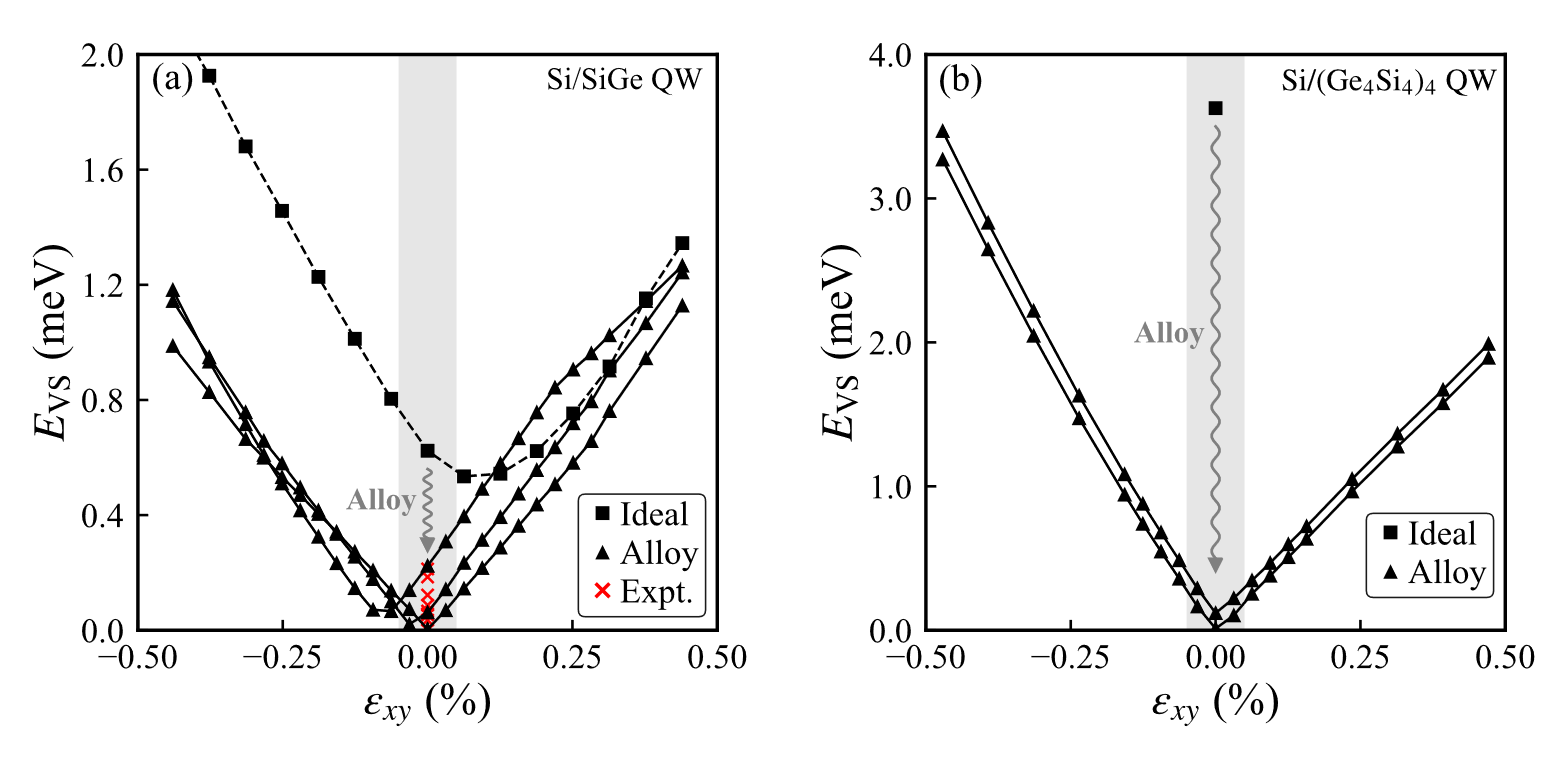}
	\caption{(a) Valley splitting energy in (001)-oriented $53~\mathrm{ML}$-$\mathrm{Si} $ QWs calculated using the atomistic SEPM. Black solid square denotes the valley splitting in the ideal $\mathrm{Si} / \mathrm{Ge}$ QW, while black solid triangles represent those in $\mathrm{Si} / \mathrm{SiGe}$ QWs with interfacial alloy disorder. The different lines correspond to three SiGe alloy configurations. Red crosses indicate experimentally reported valley splittings in $\mathrm{Si}/\mathrm{SiGe}$ QWs with interfacial alloy disorder~\cite{PhysRevApplied.13.034068, PhysRevLett.128.146802, paqueletwuetzAtomicFluctuationsLifting2022}. An electric field of $10~\mathrm{MV} / \mathrm{m}$ is applied to drive the electron to the upper interface. (b) Valley splitting energy in (001)-oriented $\mathrm{Si}(53 \mathrm{ML}) /\left(\mathrm{Ge}_4 \mathrm{Si}_4\right)_4$ QW calculated using the atomistic SEPM. An electric field of $10~\mathrm{MV} / \mathrm{m}$ is applied in the [001] direction. The square represents the ideal QW with an atomically flat interface. Scattered triangles mark the presence of interface alloy disorder, where every single $\mathrm{Si} / \mathrm{Ge}$ interface introduces a broadening width, set at 2 ML.
		\label{fig1}
	}
\end{figure}
We employ the well-established atomistic SEPM~\cite{canningParallelEmpiricalPseudopotential2000,zhang2013genetic, PhysRevB.84.121303, PhysRevLett.119.126401, PhysRevLett.104.066405, PhysRevB.92.165301, PhysRevLett.108.027401, PhysRevLett.102.056405, luoAbsenceRedshiftDirect2017} to investigate the valley splitting in Si/SiGe QWs under a bias electric field of 10 MV/m, which pushes electrons toward the upper interface. Details of the model construction and computational procedures are provided in the Methods section. As shown in Fig.~\ref{fig1}(a), our theoretical predictions are in good agreement with experimental measurements for  Si/SiGe QWs in the absence of external strain~\cite{PhysRevApplied.13.034068, PhysRevLett.128.146802, paqueletwuetzAtomicFluctuationsLifting2022}.  Specifically, an ideal 53-ML-thick Si QW embedded in a pure Ge matrix with atomically sharp interfaces yields a valley splitting $E_\mathrm{VS}$ as large as 0.7 meV---well above the 0.2 meV threshold required for reliable electron spin qubit operation~\cite{degli2024low}. In reality, however, the structural quality of Si/Ge QWs with pure Ge barriers is insufficient for qubit fabrication. Therefore, the pure Ge barrier is typically replaced with a $\mathrm{Si_{0.7}Ge_{0.3}}$ alloy to reduce lattice mismatch and improve epitaxial growth quality. As shown in Fig.~\ref{fig1}(a), this substitution dramatically decreases $E_\mathrm{VS}$ from 0.7 meV to about 0.1 meV. Even worse, different atomic configurations of the same alloy composition introduce considerable variations in valley splitting, with fluctuations exceeding 100\%---far above the uniformity required for large-scale spin qubit integration. Particularly, such inevitable atomic-scale disorder at the interface can also destroy the large valley splitting predicted for many proposed elegant QW designs~\cite{zhang2013genetic, PhysRevB.105.165308,PhysRevB.104.085406,PhysRevB.106.085304, mcjunkin2022sige}. For instance, an ordered barrier consisting of a "magic" Si/Ge atomic-layer stack, featuring a 4-ML Ge layer immediately adjacent to the Si well, was computationally designed to yield a giant valley splitting of 8.7 meV~\cite{zhang2013genetic}, while a simplified 4-period $\mathrm{Ge}_4 \mathrm{Si}_4$ superlattice barrier on one side of the Si well can also produce $E_\mathrm{VS}$ as large as 3.7 meV~\cite{PhysRevB.105.165308}. However, Fig.~\ref{fig1}(b) shows that the introduction of atom interdiffusion with a broadening of just 2 ML at each Si/Ge interface significantly reduces $E_\mathrm{VS}$ from the designed 3.7 meV to an average of merely 0.05 meV, accompanied by large configuration-dependent variations. These results again underscore the interface atomic-scale disorder being a major obstacle to achieving the large and uniform valley splitting required in real Si/SiGe QWs for scalable electron spin qubits and, ultimately, fault-tolerant quantum computing.

Here, we explore the use of shear strain as a means to overcome the disorder-induced suppression and variation of valley splitting in Si/SiGe QWs. Figure~\ref{fig1}(a) shows that for a 53-ML-thick Si QW with Si$_{0.7}$Ge$_{0.3}$ alloy barriers, the disorder-reduced valley splitting increases linearly under both tensile and compressive uniaxial strains applied along the in-plane [110] direction, with a (vanishing) minimum $E_\mathrm{VS}$ occurring slightly away from zero strain. The enhancement rate is approximately 2.4 meV per percent strain, and no sign of saturation is observed within the strain range accessible to current strained-Si CMOS technology. Specifically, a small uniaxial strain of magnitude $|\varepsilon_{xy}| = 0.3\%$---either tensile or compressive---can fully restore the disorder-reduced $E_\mathrm{VS}$ to 0.7 meV, matching the value of the ideal QW with atomically sharp interfaces. This strain-induced enhancement of $E_\mathrm{VS}$ is consistently observed across all investigated SiGe alloy configurations, with only slight shifts in the strain position of their minimum $E_\mathrm{VS}$. It is worth noting that in the ideal Si/Ge QW with atomically sharp interfaces, the minimum $E_\mathrm{VS}$ occurs at a finite tensile strain of about 0.12\% and remains as large as 0.5 meV. This behavior indicates that the nonmonotonic strain dependence of $E_\mathrm{VS}$ at small strains arises from the interplay between shear strain-induced and intrinsic valley splittings, given that the pure Ge barrier exhibits a large intrinsic valley splitting of 0.7 meV. The competition between these two contributions prevents shear strain from enhancing $E_\mathrm{VS}$ in a perfectly linear manner as conventionally expected~\cite{ohkawa1977theory}, and shifts the minimum of $E_\mathrm{VS}$ to a finite strain that depends on the alloy configuration. 

We further examine whether shear strain can restore the disorder-eliminated giant valley splitting in previously designed Si QWs~\cite{PhysRevB.105.165308}. Figure~\ref{fig1}(b) shows that tensile strain increases $E_\mathrm{VS}$ almost linearly at a rate of roughly 4 meV per percent strain, from 0.05 meV at zero strain to about 2.07 meV at $\varepsilon_{xy} = 0.5\%$ for the designed 53-ML-thick Si QW with a 4-period $\mathrm{Ge}_4 \mathrm{Si}_4$ superlattice barrier attached immediately to its upper interface.  The compressive shear strain enhances $E_\mathrm{VS}$ even more effectively as it rises from near 0.05 meV to around 3.37 meV at $\varepsilon_{xy} = -0.5\%$, corresponding to a rate of about 6.7 meV per percent strain. This asymmetric enhancement likely originates from the opposite effects of tensile and compressive uniaxial strains on the confinement energy of valley states in the Si QW, as illustrated in Fig.~\ref{figS1}, where tensile strain tends to increase the confinement energy while compressive strain tends to reduce it. A smaller confinement energy gives rise to a smaller energy separation between the valley states in the Si well and the miniband states in the $\mathrm{Ge}_4\mathrm{Si}_4$ superlattice barrier, and therefore strengthens their coupling~\cite{PhysRevB.105.165308}, which is responsible for the giant valley splitting in the designed QW~\cite{PhysRevB.105.165308}. It explains why compressive strain enhances $E_{\mathrm{VS}}$ at an even higher rate than tensile strain. Moreover, Fig.~\ref{fig1}(b) shows that strain-restored valley splitting also has a suppressed variation ratio caused by atomic disorder.

Nevertheless, despite the detrimental effect of the unavoidable atomic-scale disorder at the interfaces, our results demonstrate that the CMOS technology accessible uniaxial strain ($|\varepsilon_{xy}| \ge 0.2\%$) can effectively restore the valley splitting to beyond the 0.2 meV threshold required for spin qubit operation~\cite{degli2024low}, with a remarkable reduction in disorder-induced variability. By improving the immunity of valley splitting to the interface disorder, this approach thereby provides a practical pathway toward robust spin qubit implementation and addresses the uniform challenge in scaling up to fault-tolerant quantum computing.

%\subsection*{Theoretical model}
\subsection*{The origin of shear strain-induced valley splitting} 
\begin{figure}[t]
	\centering
	\includegraphics[width=1.0\textwidth]{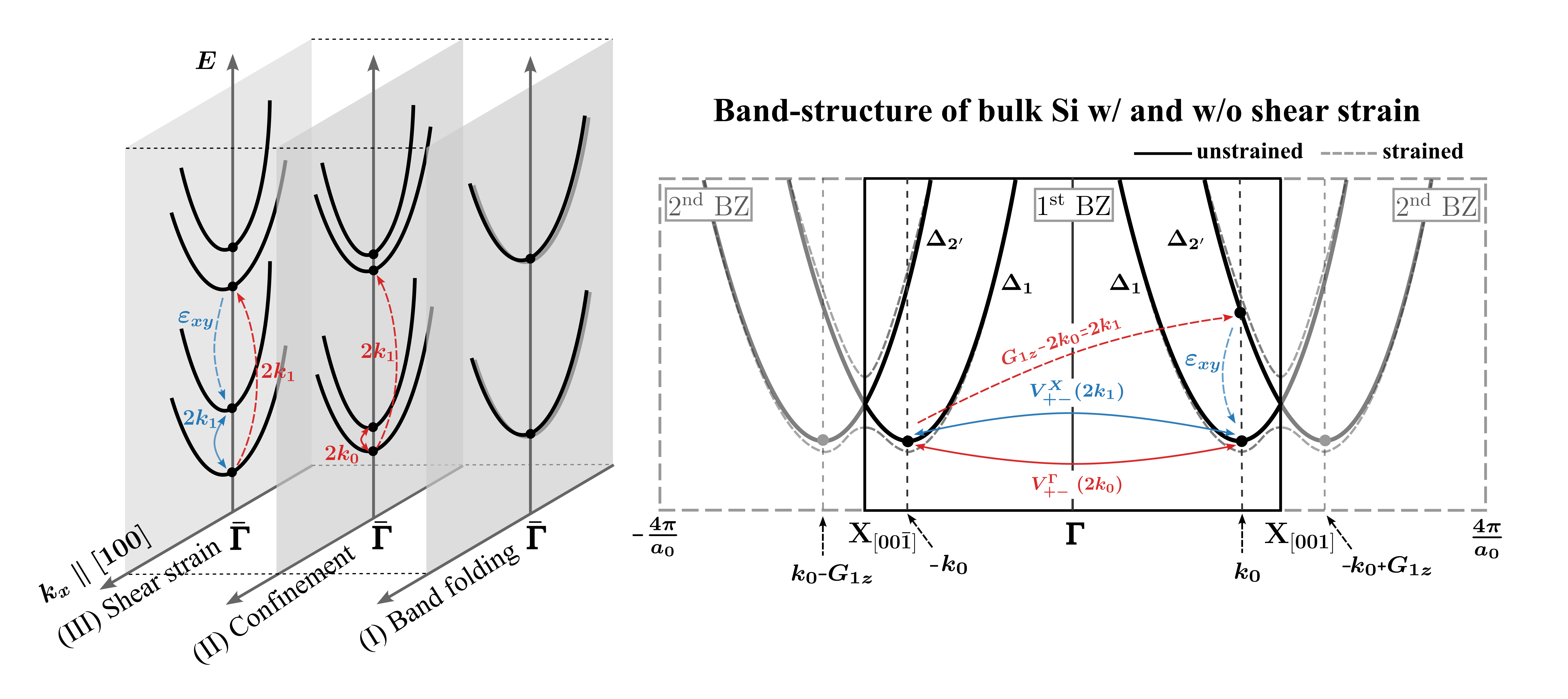}
	\caption{Schematic illustration of the valley coupling mechanism. The right panel shows the conduction band edges of bulk Si along $k_z \parallel [001]$ over two adjacent Brillouin zones, including the $\Delta_1$ and $\Delta_{2^{\prime}}$ bands. Solid and dashed curves correspond to the cases without ($\varepsilon_{xy}=0$) and with ($\varepsilon_{xy}\neq 0$) shear strain, respectively. $V_{+-}^{\Gamma}$ and $V_{+-}^X$ denote the intra-BZ and inter-BZ valley couplings. On the left, three boxes highlight successive steps that link the bulk band structure to valley splitting in Si QWs. Box-I (“band folding”) illustrates the folding of the bulk $\Delta_1$ and $\Delta_{2^{\prime}}$ bands at $\pm k_0$ onto the $\Gamma$ point, resulting in pairs of degenerate parabolic bands along $k_x \parallel [100]$. Box-II (“confinement”) shows that quantum confinement quantizes these folded bands, yielding discrete orbital states. Here, the abrupt interface potential provides Fourier components at $2k_0$ and $2k_1$, which lift the degeneracy of the lowest $\pm k_0$ states and introduce additional inter-band couplings (red); while the $2k_0$ process (solid) leads directly to valley splitting, the $2k_1$ process (dashed) leaves the lowest two levels unaffected. Finally, Box-III (“shear strain”) illustrates how shear strain activates the $2k_1$ channel, allowing it to contribute indirectly to the interaction between the lowest $\pm k_0$ states (solid blue line) and thereby enhancing the valley splitting.
	\label{fig2}
	}
\end{figure}

To elucidate the microscopic mechanism underlying the uniaxial strain-enhanced valley splitting that is protected against interfacial disorder, we extend the established theoretical framework for valley splitting in Si/Ge heterostructures~\cite{PhysRevB.84.155320} by incorporating the strain tensor. In bulk Si without strain, the conduction-band minimum (CBM) lies at $k_0 = 0.85 \times 2\pi/a_0$ along the six equivalent $\Delta$ lines, located 85\% of the way between $\Gamma$ and $X$ ($a_0$ being the lattice constant). In a Si/SiGe QW grown along the [001] direction, quantum confinement lifts the six-fold degeneracy of the $\Delta$ valleys by shifting the four in-plane valleys upward in energy relative to the two low-lying out-of-plane valleys, owing to the anisotropic effective mass. The reduced dimension along the [001] direction folds the two low-lying valley states centered at $\pm k_0$ into the reduced $\bar{\Gamma}$ point, as illustrated in Box-I in Fig~\ref{fig2}. In the absence of valley coupling, the two low-lying valley states can be expressed as the product of the bulk CBM Bloch function $\psi_{1,\pm k_0}(\boldsymbol{r})$ and the envelope function $\Psi(z)$ within the single-valley effective-mass approximation:
\begin{equation}
	\psi_{\pm}(\boldsymbol{r})=\Psi(z) \psi_{1,\pm k_0}(\boldsymbol{r})=\Psi(z) u_{1,\pm k_0}(\boldsymbol{r})e^{\pm ik_{0}z}
\end{equation}
where the periodic part $u_{1,\pm k_0}(\boldsymbol{r})$ of each Bloch function can be expanded in plane waves as
\begin{equation}
	\begin{aligned}
		u_{1, \pm k_0} (\boldsymbol{r})=\sum_{\boldsymbol{G}} c_\pm (\boldsymbol{G})e^{i{\boldsymbol{G}}\cdot \boldsymbol{r}},
	\end{aligned}
\end{equation}
with $\boldsymbol{G}$ a reciprocal lattice vector, and the corresponding coefficients $c_\pm(\boldsymbol{G})$ for Si given in Ref.~[\citen{PhysRevB.84.155320}]. In the presence of an abrupt interface potential, the above single-valley picture breaks down because the sharp spatial variation introduces high-momentum Fourier components that mediate coupling between the $\pm k_0$ valleys, thereby producing a finite valley splitting.

The Hamiltonian of a Si/SiGe QW can be described by adding the confinement potential $U(z)$ and an external electric field $F$ to the bulk Si Hamiltonian $H_0$,
\begin{equation}
	H=H_0+U_{\mathrm{total}}(z)=H_0+U(z)-\frac{eF}{\epsilon}z,
\end{equation}
where the confinement potential is defined as $U(z)=0$ for $z_-<z<z_+$ and $U(z)=U_0$ otherwise ($U_0=150$ meV for a Si$_{0.7}$Ge$_{0.3}$ barrier). Projecting the total Hamiltonian onto the $\psi_\pm(\boldsymbol{r})$ basis yields an effective two-level system,
\begin{equation}
	\bar{H}=
	\begin{pmatrix}
		E_{0} & U_\mathrm{V}  \\
		U_\mathrm{V}^{*} & E_{0}
	\end{pmatrix},
\end{equation}
where $U_\mathrm{V}$ couples two low-lying valleys, giving rise to a valley splitting energy $E_\mathrm{VS}=2|U_\mathrm{V}|$. By expanding the Bloch functions $\psi_\pm(\boldsymbol{r})$ in plane-wave basis, we obtain
\begin{equation}
	U_V=\langle \psi_{+}|U_\mathrm{total}(z)|\psi_{-}\rangle=\sum_{{\bf G}}B_{{\Delta_1},-k_0;\Delta_1,k_0}({\bf G}) \delta_{G_x,0}\delta_{G_y,0} \tilde{U}_\psi(-2k_0+G_z),
	\label{eq.coupling}
\end{equation}
where $\tilde{U}_\psi(q_z)=\int_{-\infty}^{\infty} U_\mathrm{total}(z) |\Psi(z)|^2 e^{-iq_z z}dz$ is the Fourier transform of the total potential weighted by the envelope function, and $B_{{\Delta_1},-k_0;\Delta_1,k_0}({\bf G})=\sum_{{\bf G}^\prime} c_{+}^*({\bf G}^\prime)c_{-}({\bf G}^{\prime}+{\bf G})$ represents the overlap between plane-wave components differing by a reciprocal lattice vector ${\bf G}$. Equation~\ref{eq.coupling} indicates that two low-lying valley states are coupled through a summation over the Fourier components involving a momentum transfer of $- 2k_0 + G_z$, modulated by the overlap coefficient $B_{{\Delta_1},-k_0;\Delta_1,k_0}({\bf G})$~\cite{PhysRevB.84.155320, PhysRevB.106.085304, klymenko2014electronic, resta1977note}. This overlap encods the crystal symmetry selection rules that determine which ${\bf G}$ components effectively contribute to valley coupling. For Si/SiGe QWs without shear strain, the dominant term of the valley coupling arises from the direct overlap at ${\bf G=0}$ with $B_{{\Delta_1},-k_0;\Delta_1,k_0}(0) \approx -0.2607$, which involves a large momentum transfer of $2k_0$ through the $\Gamma$ point, as shown in Fig~\ref{fig2}, and is refered to as intra-band ($\Delta_1$-$\Delta_1$) coupling. It constitutes the intrinsic valley coupling responsible for the finite valley splitting typically observed in Si/SiGe QWs with atomically sharp interfaces.

Note that the interface-induced breaking of translational symmetry also enables coupling between these two low-lying valley states and the higher conduction-band states folded to the reduced $\bar \Gamma$ point. In the diamond lattice,  the second-lowest conduction band is denoted as $\Delta_{2^\prime}$ along the  $\Delta$ direciton and is degenerated with the lowest $\Delta_1$ band at each $X$ point, as enforced by the glide-reflection symmetry of the $O_h$ group---a feature often referred to as the “sticking-together” behavior~\cite{PhysRev.138.A225,jones1960theory}. As illustrated in Box-II of Fig.~\ref{fig2}, the total potential $U_{\mathrm{total}}$ can mediate coupling between the $\Delta_1$ band $k_0$ ($-k_0$) valley and the $\Delta_{2^\prime}$ band at the opposite $-k_0$ ($k_0$) point in terms of bulk Brillouin zone (BZ). The matrix element of this inter-band ($\Delta_1$-$\Delta_{2^\prime}$) coupling has the same form as the intra-band ($\Delta_1$-$\Delta_1$) coupling in Eq.~\ref{eq.coupling}, except that the overlap coefficient is replaced by $B_{{\Delta_1},-k_0;\Delta_{2^\prime},k_0}({\bf G})$. Since the $\Delta_{2^\prime}$ band can be regarded as the same band as the $\Delta_1$ band in the context of extended BZ, shifted by a reciprocal lattice vector ${\bf G}_1=(0,0,\frac{4\pi}{a_0})$ into the first BZ, their periodic parts are approximately related by $u_{\Delta_{2^\prime},{\pm {\bf k}_0}}({\bf r})\approx u_{\Delta_1,{\mp {\bf k}_0}}({\bf r})e^{\mp i{\bf G}_1\cdot {\bf r}}$. As a result, the direct overlap term $B_{{\Delta_1},-{\bf k}_0;\Delta_{2^\prime},{\bf k}_0}(0)\approx 0$ vanishes, whereas the Umklapp process $B_{{\Delta_1},-{\bf k}_0;\Delta_{2^\prime},{\bf k}_0}({\bf G}_1)\approx 1$ dominates the inter-band ($\Delta_1$-$\Delta_{2^\prime}$) coupling. Thus, only the Fourier component at $G_{1z}-2k_0=2k_1$ contributes to this inter-band coupling, distinct from the conventional intra-band coupling. It is worth noting that this term does not directly affect the valley splitting, as the coupling does not modify the interaction between the two low-lying valleys but merely introduces an overall energy shift.

However, this inter-band ($\Delta_1$-$\Delta_{2^\prime}$) coupling plays a critical role in uniaxial strain-induced valley splitting. Applying a shear strain $\varepsilon_{xy}$ to lower the crystal symmetry from $O_h$ to $D_{2h}$ lifts the degeneracy by opening an anticrossing gap between the $\Delta_1$ and $\Delta_{2'}$ bands at the $X$ point (Fig.~\ref{fig2}). Using the periodic parts of these two bands as basis functions, the low-energy states near the BZ boundary can be described by a two-band $k\cdot p$ Hamiltonian~\cite{PhysRev.138.A225,PhysRev.142.530}:
\begin{equation}
	H^s_0=\left(E_X+\frac{\hbar^2 k_z^2}{2 m_l}\right) \boldsymbol{{1}}+D \varepsilon_{x y} \boldsymbol{\sigma_x}+\frac{\hbar^2 k_1 k_z}{m_l} \boldsymbol{\sigma_z},
\end{equation}
where $k_z$ is measured from the $X$ point, $k_1 = 0.15\times2\pi/a_0$ gives the valley position relative to $X$, $m_l=0.19m_0$ is the longitudinal effective mass, and $D=5.7$ eV is the deformation potential for uniaxial shear strain. It reads that the shear strain couples the originally independent $\Delta_1$ and $\Delta_{2'}$ bands, producing an anticrossing gap $2D\varepsilon_{xy}$ at the $X$ point. This coupling modifies the periodic part of the valley Bloch function by mixing $\Delta_1$ and $\Delta_{2'}$ bands: 
\begin{equation}
	%	\small
	u^\prime_{\pm k_0}(\boldsymbol{r}) = \cos{\left(\frac{\phi}{2}\right)} u_{1, \pm k_0} (\boldsymbol{r})-\text{sgn}(\varepsilon_{x y}) \sin{\left(\frac{\phi}{2}\right)}  u_{2, \pm k_0}(\boldsymbol{r}),
\end{equation}
where $\phi=\arctan \left(  D\varepsilon_{x y}/ \hbar^2 k_1^2  \right)$, and $u_{1, \pm k_0}(\boldsymbol{r})$ and $u_{2,\pm k_0}(\boldsymbol{r})$ are the periodic parts of the $\Delta_1$ and $\Delta_{2^{\prime}}$ Bloch functions at the $\pm k_0$ points in unstrained Si, respectively. Although the shear strain-induced coupling shifts the valley minima from $\pm k_0$ in unstrained Si to $\pm k_0^\prime=\pm2\pi/a_0\mp \sqrt{ k_{1}^2-(m_l D\varepsilon_{xy}/\hbar^2 k_1)^{2} }$, here, we still approximate them as $\pm k_0$ since the applied strain is small. Through shear strain, the $\Delta_1$ and $\Delta_{2'}$ bands are non-perturbatively mixed at each valley minimum, thereby activating an inter-band ($\Delta_1-\Delta_{2'}$) coupling channel with momentum transfer $2k_1$ that mediates interaction between the $\pm k_0$ valleys (Box-III, Fig.~\ref{fig2}).

Building upon the above framework, we can derive an analytical form for the valley coupling under uniaxial shear strain by substituting the strain-modified Hamiltonian and Bloch functions into Eq.~\ref{eq.coupling}. Because the applied electric field $F$ confines electrons near either the upper or lower interface ($z_+$ or $z_-$), only one interface contribution is considered, while higher-order strain terms are neglected [see Support Materials (SM) for details]. The resulting coupling is written as
\begin{equation}
	\begin{aligned}
		U_\mathrm{V}=&\left|\Psi\left(z_i\right) \right|^2 \left[\cos^2(\frac{\phi}{2})  \frac{\Sigma_0 U_0}{i2k_0}e^{-2ik_0z_i} \right.  \left. +\text{sgn}(\varepsilon_{x y}) \sin\phi \frac{\Sigma_1 U_0}{i2k_1}e^{2ik_1 z_i}  \right],\\
	\end{aligned}
	\label{eq.coupling_total}
\end{equation}
where $z_i=z_+$ for a positive electric field and $z_-$ for a negative one. The constants $\Sigma_0$ and $\Sigma_1$ denote the overlap coefficients $B_{1,-k_0;1,k_0}(0)$ and $B_{1,-k_0;2,k_0}(G_1)$, respectively. The first term corresponds to the intrinsic (strain-free) valley coupling $U_\mathrm{V}$ widely described in the literature~\cite{ohkawa1977theory, PhysRevB.75.115318, PhysRevB.84.155320}. It arises from the intra-band $2k_0$ coupling channel through the $\Gamma$ point of the $1^{\text{st}}$ BZ and is therefore referred to as the intra-BZ valley coupling. In contrast, the second term describes an indirect valley coupling mediated by the inter-band $2k_1$ coupling channel, which becomes active only in the presence of shear strain. As this $2k_1$ process can formally be viewed as coupling valleys located in adjacent BZs, it is thus identified as the inter-BZ valley coupling. Since $\sin\phi \propto |\varepsilon_{xy}|$, its strength grows linearly with the applied shear strain and can even exceed the intrinsic intra-BZ contribution at sufficiently large strain~\cite{noborisakaValleySplittingExtended2024}.

%Since $\sin \phi \propto |\varepsilon_{xy}|$, this term grows linearly with shear strain. Moreover, given that the minima of the shear-strain-coupled $\Delta_1$ and $\Delta_{2'}$ bands are located near one of the X points but in the first BZ and second BZ, respectively, we thus refer to the second term as the inter-BZ valley coupling [origin and green lines in the Fig.~\ref{fig2}]. At sufficiently large shear strain, this inter-BZ contribution can even dominate over the intrinsic intra-BZ coupling~\cite{noborisakaValleySplittingExtended2024}. 

Based on Eq.~\ref{eq.coupling_total}, the valley splitting energy can be expressed as:
\begin{equation}
	\begin{aligned}
		E_\mathrm{VS}=&2 \left|\cos^2 \left(\frac{\phi}{2}\right)  \frac{\left|\Sigma_0\right|U_0}{2k_0}e^{i\left[\pi - 4\pi   z_i/a_0\right]}  \right. + \left. \text{sgn}(\varepsilon_{x y}) \sin\phi \frac{\left|\Sigma_1\right|U_0}{2k_1}   \right| \cdot \left|\Psi\left(z_i\right) \right|^2 		\equiv 2\left|V_{+-}^\Gamma e^{i\theta} +V_{+-}^X \right|.
	\end{aligned}
	\label{eq.vs}
\end{equation}
Here, $V_{+-}^\Gamma$ and $V_{+-}^X$ represent the contributions to valley splitting arising from the intra-BZ valley coupling and inter-BZ valley coupling, respectively. Their combined effect on the total valley splitting energy is modulated by the interface position $z_i$ (or equivalently, the QW thickness), through a relative phase angle $\theta=\pi - 4\pi z_i/a_0$. This relation provides a microscopic basis for understanding shear strain-induced valley splitting $V_{+-}^X$ in QWs, as well as its interference with the intrinsic valley splitting $V_{+-}^\Gamma$ that exists in the absence of strain. Specifically, in the condition $V_{+-}^X=-V_{+-}^\Gamma \cos \theta$, the total valley splitting $E_\mathrm{VS}$ is minimized, explaining the strain-induced suppression of valley splitting and the minimum $E_\mathrm{VS}$ occurred at nonzero strain, as observed in Fig.~\ref{fig1}(a). 

To validate our valley splitting model, we performed atomistic simulations of ideal Si/Ge QWs with varying thicknesses and opposite electric field orientations [Full simulation details, and corresponding figures are provided in the SM]. The results reproduce the predicted periodic variations of the relative phase angle $\theta$ with QW thickness and reveal the expected sign reversal of $\theta$ under electric field inversion for odd-monolayer wells, but not for even-monolayer ones. These findings highlight how atomic-scale variations in interface positioning lead to interference between inter- and intra-BZ valley coupling, thereby shaping the strain response of valley splitting. 

\begin{figure}[t]
	\centering
	\includegraphics[width=0.48\textwidth]{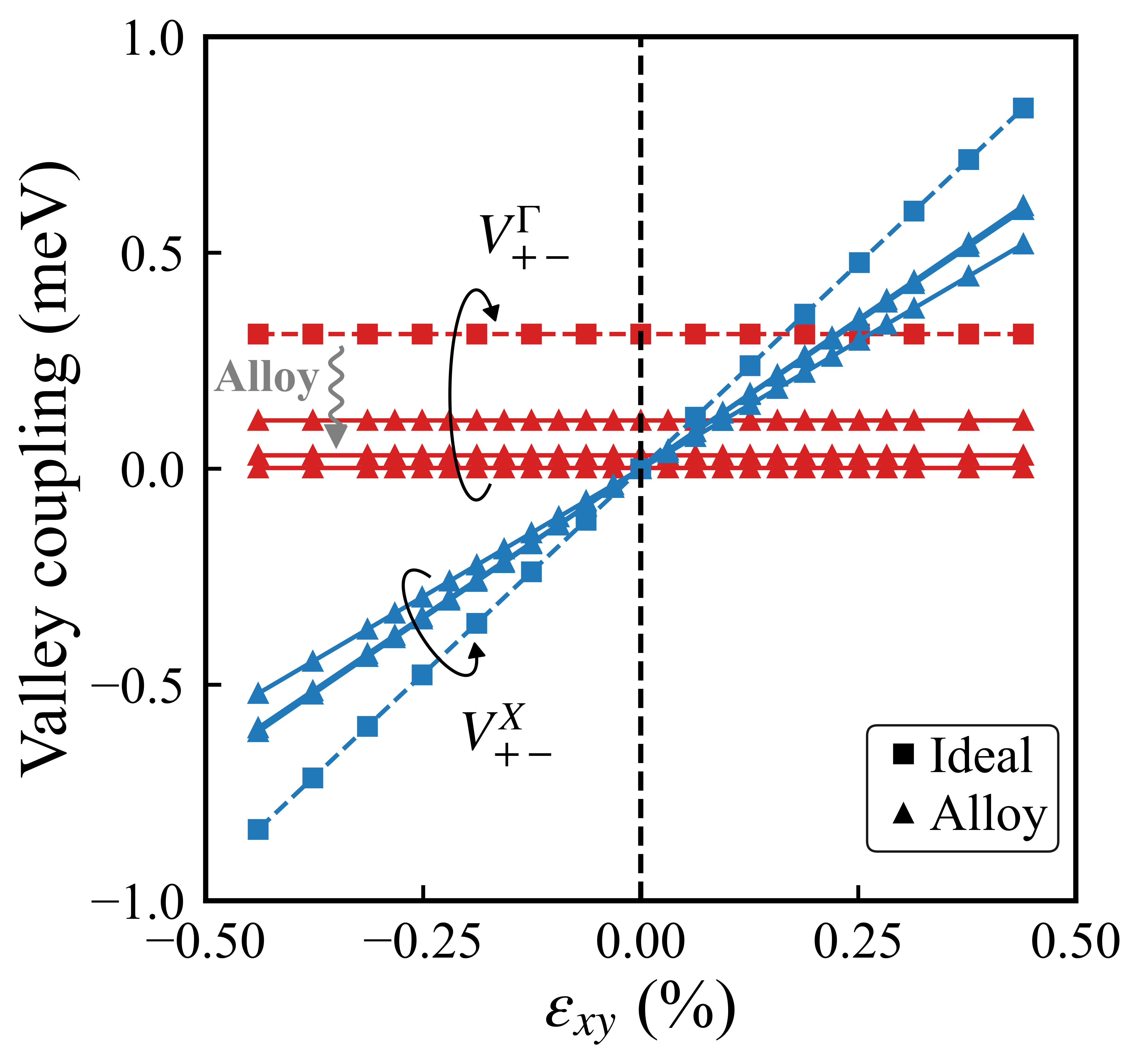}
	\caption{Dependence of inter-BZ valley coupling $V_{+-}^X$ (blue) and intra-BZ valley couplings $V_{+-}^{\Gamma}$ (red) on shear strain in a 53 ML-thick Si QW. Solid squares represent the scenario of an ideal Si QW, and solid triangles correspond to Si QWs incorporating various interfacial alloy configurations, corresponding to the representative cases in Fig.~1(a). $V_{+-}^X$ is extracted using Eq.~\ref{eq.vs}, with atomistically calculated $E_{\mathrm{VS}}$ as input. $V_{+-}^{\Gamma}$ remains constant with strain, corresponding to $E_{\mathrm{VS}} / 2$ in the absence of shear strain.
		\label{fig3}
	}
\end{figure}

By fitting atomistic SEPM-calculated valley splitting $E_{\mathrm{VS}}$ versus shear strain to Eq.~\ref{eq.vs}, we can now extract the parameters $V_{+-}^{\mathrm{\Gamma}}$, $V_{+-}^{\mathrm{X}}$, and the relative phase angle $\theta$ for QWs with both ideal and alloy barriers. In this model, $V_{+-}^{\mathrm{\Gamma}}$ is treated as a fixed positive constant, defined as half the value of $E_{\mathrm{VS}}$ at $\varepsilon_{xy} = 0$, while $V_{+-}^{\mathrm{X}}$ is assumed to vary linearly with $\varepsilon_{xy}$. As shown in Fig.~\ref{fig3}, the fitted results indicate that the well-known suppression of $E_\mathrm{VS}$ due to interfacial disorder in Si QWs originates solely from the reduction of the intrinsic valley splitting term $V_{+-}^\Gamma$, whereas the shear strain-induced contribution $V_{+-}^X$ remains largely unaffected. This critical distinction indicates that while interface roughness has long been regarded as the primary obstacle to achieving high and uniform valley splitting, shear strain provides a pathway that circumvents this limitation. In the following section, we delve into the physical origin of this immunity, explaining how shear strain reshapes the valley coupling to make them resilient against atomic-scale disorder.

\subsection*{Why strain-induced valley splitting is protected against interface disorder}

In contrast to ideal pure-Ge barriers, alloy fluctuations in realistic Si/SiGe QWs---recognized as the primary source of disorder---affect the electronic states mainly in two ways: by altering the strength of the localized confinement potential and by causing fluctuations in the interface position, which results in an effective interface width rather than a sharp boundary. While the first effect has a minimal impact on valley splitting due to averaging along the growth direction, the second effect, interface broadening, plays a critical role in understanding the detrimental impact of disorder. In atomistic terms, alloy disorder at the interface gives rise to a complex landscape of step-like features, where local variations in composition cause monolayer-scale fluctuations in the interface position~\cite{10.1063/1.2591432, Lima_2023, PhysRevMaterials.8.036202}. These fluctuations can be viewed as a superposition of multiple atomic steps randomly distributed along the interface. To capture the essence of this interface broadening without resorting to fully disordered simulations, we adopt a simplified single-step model to provide a tractable and physically meaningful approximation, which both represents a minimal unit of alloy-induced interface fluctuation and constitutes another form of interface disorder~\cite{PhysRevResearch.2.043180, PhysRevB.100.125309}.

\begin{figure}[t]
	\centering
	\includegraphics[width=0.96\textwidth]{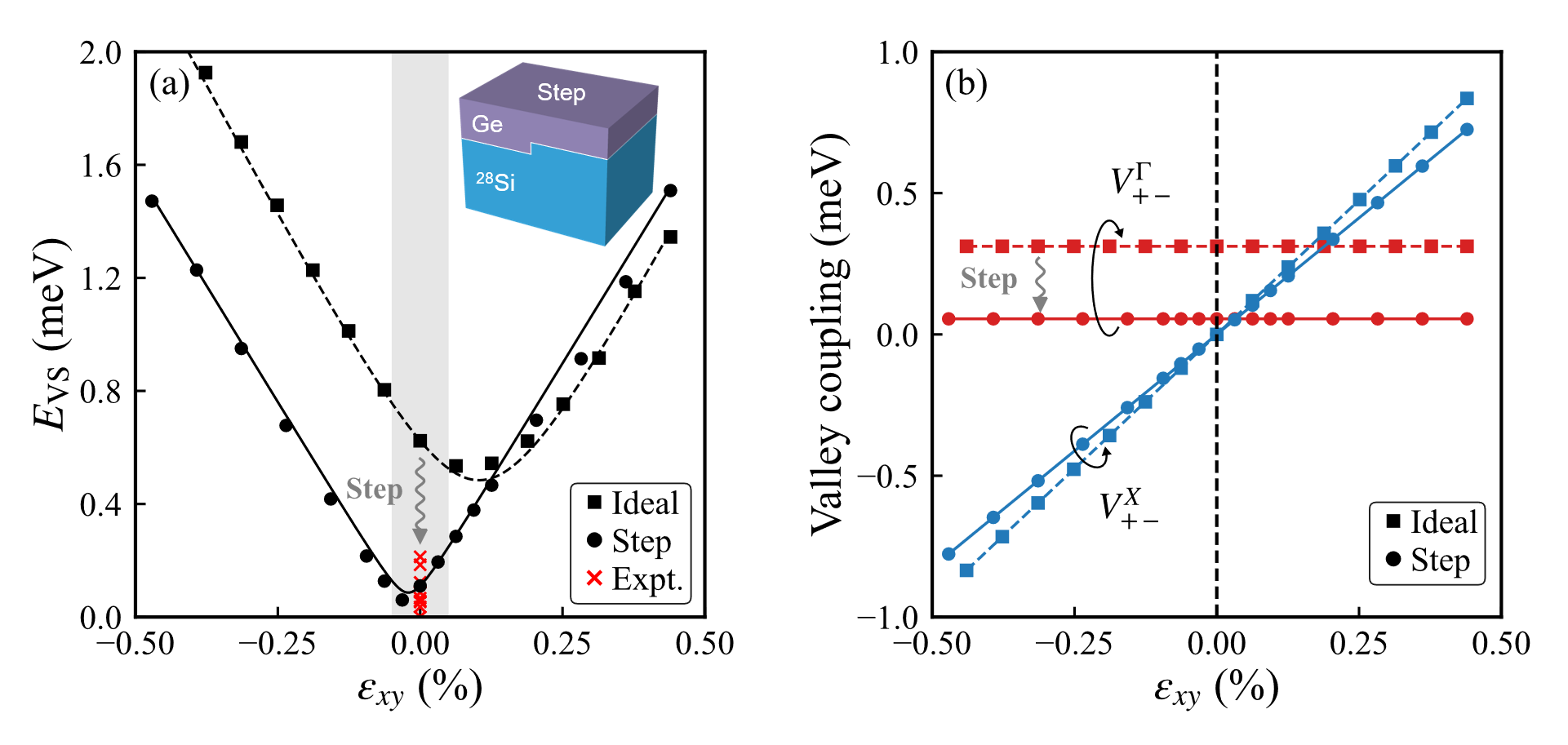}
	\caption{(a) Valley splitting energy in (001)-oriented $53~\mathrm{ML}$-$\mathrm{Si}/\mathrm{Ge}$ QWs calculated using the atomistic SEPM. Black solid squares represent results for $\mathrm{Si} / \mathrm{Ge}$ QWs with atomically flat interfaces, while black solid circles depict the case with interface steps. Red crosses indicate experimentally reported valley splittings in $\mathrm{Si}/\mathrm{SiGe}$ QWs with interfacial alloy disorder~\cite{PhysRevApplied.13.034068, PhysRevLett.128.146802, paqueletwuetzAtomicFluctuationsLifting2022}. The solid/dashed lines represent fits based on Eq.~\ref{eq.vs}. The inset illustrates the configuration of a single-step interface. (b) Dependence of inter-BZ valley coupling $V_{+-}^X$ (blue) and intra-BZ valley couplings $V_{+-}^{\Gamma}$ (red) on shear strain in a 53 ML-thick Si QW. Solid squares represent the scenario of an ideal Si QW, and solid circles correspond to Si QWs with the interface step. A uniform electric field of $10~\mathrm{MV} / \mathrm{m}$ is applied during the atomistic SEPM simulation.
		\label{fig4}
	}
\end{figure}

In the single-step model, we focus on the Si/Ge QWs rather than Si/SiGe QWs in our atomistic simulations. Specifically, we construct a 1-ML-high atomic step edge, oriented along the [010] direction, positioned at the center of the [100] direction on the interface plane (001). This configuration serves as a minimal yet representative model that captures the essential features of alloy-induced interface broadening. Its simplicity enables a transparent analysis of the mechanism underlying the responses of both valley couplings to interfacial disorder. As shown in Fig.~\ref{fig4}(a), the simulation results for this single-step model effectively reproduce the suppression of valley splitting due to the alloy disorder, which reduces it to approximately 0.11 meV, as well as a shift in the strain condition where the valley splitting minimum occurs, from tensile strain in the ideal QW to compressive strain in the disordered case [see Fig.~\ref{fig1}(a)]. These results demonstrate the effectiveness of the single-step model in capturing the disorder-induced suppression mechanism.

We further examine the sensitivity of the intra-BZ and inter-BZ valley couplings to the presence of interfacial steps by fitting the atomistically predicted $E_{\mathrm{VS}}$ [shown in Fig.~\ref{fig4}(a)] using Eq.~\ref{eq.vs}. Figure~\ref{fig4}(b) shows that the extracted value of $V_{+-}^{\mathrm{\Gamma}}$ for an ideal 53-ML-thick (001)-oriented Si/Ge QW is 0.35 meV, which is reduced by an order of magnitude to 0.055 meV when interfacial steps are introduced. In contrast, the shear strain-induced $V_{+-}^X$  exhibits an almost unchanged rate of increase with strain, regardless of the presence of the step. This sharp contrast in the responses of the two types of valley coupling to the interfacial step stems from their distinct phase angles, which originate from the differing momentum transfers involved in scatting across valleys in the BZs, as schematically illustrated in Fig.~\ref{fig2}: For a 1-ML height step that bisects the plane of the QW upper interface, the contributions to valley coupling from either side of the step have the same magnitude but significantly different phase angles. This phase mismatch alters the net valley splitting energy according to Eq.~\ref{eq.vs}:
\begin{equation}
	E_\mathrm{VS}=2\left| V_{+-}^{\Gamma 0} e^{i\theta} \left(\frac{1+e^{-ik_0 a_0/2}}{2} \right)  +   V_{+-}^{X 0}  \left( \frac{1+e^{ik_1 a_0/2}}{2} \right)  \right|=2 \left| \eta^\Gamma \cdot V_{+-}^{\Gamma 0} e^{i(\theta-\frac{\pi}{2})}  +  \eta^X \cdot V_{+-}^{X 0}   \right|
\end{equation}
where $\eta^\Gamma=\left|\left(1+e^{-ik_0 a_0/2}\right)/2\right|$ and  $ \eta^X=\left|\left(1+e^{ik_1 a_0/2}\right)/2\right|$ are the phase-related reduction factors for the intra- and inter-BZ valley coupling components $V_{+-}^{\Gamma 0}$ and $V_{+-}^{X 0}$ of an ideal QW, respectively, caused by the presence of a single interface step. It follows that a larger phase angle difference leads to greater sensitivity of the valley coupling magnitude to variations in the interface position caused by the step. Specifically, the phase angle difference for the intra-BZ valley coupling is $-k_0 a_0/2=-0.85\pi$, which results in a strong suppression of the coupling strength to only $\eta^\Gamma$=23.3\% of the perface interface value $V_{+-}^{\Gamma 0}$. In contrast, the inter-BZ coupling experiences a much smaller phase difference of $k_1 a_0/2=0.15\pi$, retaining $\eta^X = 97.23\%$ of the ideal value $V_{+-}^{X 0}$. Since the inter-BZ valley coupling is entirely induced by shear strain, this analysis reveals the atomistic origin of the robustness of shear strain-enhanced valley splitting against interfacial steps and alloy disorder in Si QWs---namely, the smaller phase difference arising from the reduced momentum transfer makes it less prone to cancellation by the fluctuation of interface positions.

Moreover, the presence of an 1-ML height step shifts the shear strain condition for the valley splitting minimum from $V_{+-}^{X0}=-V_{+-}^{\Gamma 0} \cos \theta$ in ideal QWs to $V_{+-}^{X0}=-V_{+-}^{\Gamma 0} \sin \theta \cdot \eta^\Gamma/\eta^X$, thereby altering both its sign and magnitude. This explains the shift in the strain corresponding to the valley splitting minimum from $\varepsilon_{xy}=0.28$\% to $\varepsilon_{xy}=-0.03$\% in the 53-ML thick Si/Ge QW upon introducing the interface step, as shown in Fig.~\ref{fig4}(a). This simple analysis also qualitatively reproduces the behavior observed in Fig.\ref{fig1}(a) for disordered interfaces. Therefore, in realistic QWs, interfacial steps and alloy disorder can partially mitigate the suppressive effect of shear strain on valley splitting.

%To more comprehensively approximate the impact of alloy disorder, we further analyzed the model-predicted influence of step position and height on the intra- and inter-VO coupling amplitudes, as shown in Fig.\ref{fig4}(c). It is evident that for a given step configuration, the reduction in $V_{+-}^{X}$ is consistently smaller than or equal to that in $V_{+-}^{\Gamma}$. This trend is further supported by Fig.\ref{fig4}(d), where the fitted results of valley splitting under different alloy configurations (as in Fig.~\ref{fig1}(a)) confirm the robustness of this relationship.

\section*{Discussion}

In summary, we demonstrate that shear strain in Si/SiGe QWs provides an effective approach to mitigating the significant suppression of valley splitting induced by interfacial disorder. Using an envelope function model that incorporates both intra-BZ and inter-BZ valley coupling, we reveal the intricate dependence of valley splitting on shear strain, a relationship further validated by atomistic simulations. On one hand, the small momentum transfer (phase angle) associated with shear strain-induced inter-BZ valley coupling renders its contribution to $E_\mathrm{VS}$ largely insensitive to interface steps and alloy disorder. Atomistic simulations show that a modest shear strain of just 0.2\% is sufficient to stabilize valley splitting for spin qubit applications, preventing further suppression due to disorder. This highlights the role of shear strain in enhancing the uniformity of valley splitting in Si spin qubit devices. On the other hand, the interplay between intra-BZ and inter-BZ valley coupling can initially suppress valley splitting under small compressive or tensile shear strain, which may originate from metallic gates, dislocations, or dielectric layers~\cite{PhysRevApplied.20.024056}. This mechanism provides an additional pathway that may lead to reduced and less uniform valley splitting in existing devices.

\section*{Methods}
We calculate the electronic structures of $\mathrm{Si} / \mathrm{SiGe}$ QWs by directly diagonalization of the system Hamiltonian $- \frac{\nabla^2}{2}+V(\mathbf{r})$ within a plane-wave basis~\cite{PhysRevB.51.17398}, utilizing the folded spectrum method~\cite{10.1063/1.466486}. The potential $V(\mathbf{r})$ comprises two components: (i) a local part, $\hat{v}_\alpha^{\text {loc }}(\mathbf{r})$, treated in reciprocal space~\cite{PhysRevB.48.11204}, and (ii) a nonlocal spin-orbit interaction part, $\hat{v}_\alpha^{\text {nonloc }}(\mathbf{r})$, implemented via the Kleinmen-Bylander formalism~\cite{PhysRevLett.48.1425}. The local pseudopotential is constructed as a superposition of screened pseudopotentials $v_\alpha(r)$ of the constituent atom~\cite{PhysRevB.51.17398, PhysRevB.59.5678},
\begin{equation}
	V(\mathbf{r})=\sum_n \sum_\alpha \hat{v}_\alpha\left(\mathbf{r}-\mathbf{R}_n-\mathbf{d}_\alpha\right),
\end{equation}
where $\hat{v}_\alpha\left(\mathbf{r}-\mathbf{R}_n-\mathbf{d}_\alpha\right)$ represents the screened pseudopotential of atom type $\alpha$ at site $\mathbf{d}_\alpha$ in the $n$th primary cell $\mathbf{R}_n$. Our approach employs a supercell combined with periodic boundary conditions. The screened pseudopotentials, $\left\{\hat{v}_\alpha\right\}$, are fitted~\cite{PhysRevB.51.17398} to mitigate the ``LDA error" in the bulk crystal and to accurately reproduce the band gaps throughout the zone, as well as the electron and hole effective-mass tensors, valence band and conduction band offsets between well and barrier materials, spin-orbit splittings, and GW spin-splittings in bulk materials~\cite{PhysRevLett.102.056405}.

To explore the impact of interface alloying, we consider practical device configurations, specifically $\mathrm{Si}/ \mathrm{Si}_{0.7} \mathrm{Ge}_{0.3}~\mathrm{QW}$ on a SiGe buffer layer. Multiple $\mathrm{Si} / \mathrm{Si}_{0.7} \mathrm{Ge}_{0.3}~\mathrm{QW}$ structures are generated using direct sampling methods, with the $\mathrm{Si}_{0.7} \mathrm{Ge}_{0.3}$ random alloy segment comprising 3600 atoms. The in-plane lattice constant of the supercell is set to match the bulk $\mathrm{Si}_{0.7} \mathrm{Ge}_{0.3}$ lattice constant, according to Vegard's law. Subsequently, we ascertain the optimized lattice constant in the vertical direction and atomic equilibrium positions by minimizing the strain energy of the entire supercell. This optimization process employs the atomistic valence force field method~\cite{10.1063/1.366631, PhysRevB.62.12963}.

\bibliography{ref}

\section*{Acknowledgements (not compulsory)}

The work was supported by the National Science Fund for Distinguished Young Scholars under Grant No. 11925407, CAS Project for Young Scientists in Basic Research under Grant No. YSBR-026,  the National Natural Science Foundation of China under Grant Nos. 12374078, and 12504094, and the Postdoctoral Fellowship Program of China Postdoctoral Science Foundation under Grant No. GZC20252226.

\section*{Author contributions statement}

Y.L. developed the model and interpreted the calculations. G.W. performed the calculations. J.-W.L. conceived the research project. S.G., J.-W.L. and S.S.L supervised the project, contributed to the discussion, and reviewed the results. All authors reviewed the manuscript.

\section*{Additional information}

Supplementary Information accompanies this paper at xxx.

\newpage

\begin{center}
	{\large \textbf{Supplemental Material: \\ Protected valley splitting against interface disorder toward scalable silicon electron spin qubits}}\\[0.2cm]
	Yang Liu$^{1,+}$, Gang Wang$^{2,+}$, Shan Guan$^{1}$, Jun-Wei Luo$^{1,*}$, and Shu-Shen Li$^{1}$\\[0.1cm]
	{\itshape $^1$State Key Laboratory of Semiconductor Physics and Chip Technologies, Institute of Semiconductors, Chinese Academy of Sciences, Beijing 100083, China}\\
	{\itshape $^2$School of Physics, Ningxia University, Yinchuan 750021, China}\\
\end{center}

%\begin{center}
%	\textbf{\large {Supplemental Material: \\ Shear Strain Control of Valley Splitting in Si/SiGe Heterostructures for Enhanced Interface Disorder Resilience}\\[.2cm]}
%	Yang Liu$^{1,2}$, Gang Wang$^{1,2}$, Shan Guan$^{1}$, and Jun-Wei Luo$^{1,2,*}$\\[.1cm]
%	{\itshape $^1$State Key Laboratory of Superlattices and Microstructures, Institute of Semiconductors, Chinese Academy of Sciences, Beijing 100083, China}\\
%	{\itshape $^2$Center of Materials Science and Optoelectronics Engineering, University of Chinese Academy of Sciences, Beijing 100049, China}\\
%\end{center}

\setcounter{equation}{0}
\setcounter{figure}{0}
\setcounter{table}{0}
\setcounter{section}{0}

\renewcommand{\theequation}{S\arabic{equation}}
\renewcommand{\thefigure}{S\arabic{figure}}
\renewcommand{\thesection}{S\arabic{section}}

\section*{Abstract}
In the Supplemental Material, we provide details on the basis function approximation for $u_{2,\pm k_0}(\boldsymbol{r})$ (S1) and the derivation of valley coupling (Eq.~\ref{eq.coupling_total}, S2). In S3, we report the calculated wave functions of Si/Ge QWs under different shear strains. Section 4 presents atomistic simulations of ideal Si/Ge quantum wells (QWs) with varying thicknesses and opposite electric field orientations, which validate the proposed valley splitting model by analyzing the relative phase angle $\theta$ and its impact on strain-dependent behavior. In S5, we present the atomistic results for the configuration with doubled step area density.

\section{Basis function approximation}
Owing to the diamond structure characteristic of bulk silicon, the $\Delta_1$ and $\Delta_{2^{\prime}}$ bands along the (001)-direction become degenerate at the $X$ point and possess a nonzero slope within the first Brillouin zone, where $k_z$ lies within the range of $\left(-2\pi/a_0, 2\pi/a_0\right]$. Additionally, it should be noted that the $\Delta_{2^{\prime}}$ band is, in fact, the folding band of the $\Delta_1$ band in the second Brillouin zone, with $k_z$ falling within the interval of $\left(-4\pi/a_0, 4\pi/a_0\right]$. Consequently, we are able to analyze the correlation between the wave functions of these two bands through the transformation between the extended Brillouin zone and the reduced Brillouin zone. 

We can define the wave function for the valley position of the $\Delta_1$ band in the extended Brillouin zone as $\phi_{\pm k_0}(\boldsymbol{r})=v_{\pm k_0}(\boldsymbol{r})e^{\pm ik_0 \boldsymbol{r}}$, and for the valley positions of the $\Delta_1$ and $\Delta_2^\prime$ bands in the reduced Brillouin zone as $\psi_{1,\pm k_0}(\boldsymbol{r})=u_{1,\pm k_0}(\boldsymbol{r})e^{\pm ik_0 \boldsymbol{r}}$ and $\psi_{2,\pm k_0}(\boldsymbol{r})=u_{2,\pm k_0}(\boldsymbol{r})e^{\pm ik_0 \boldsymbol{r}}$, respectively. Through the process of band folding, we can observe that $\psi_{1,\pm k_0}(\boldsymbol{r})=\phi_{\pm k_0}(\boldsymbol{r})$ and $\psi_{2,\pm k_0}(\boldsymbol{r})=\phi_{\pm (k_0-\boldsymbol{G}_1)}(\boldsymbol{r})$, where $\boldsymbol{G}_1=(0,0,4\pi/a_0)$. 

In the absence of strain, the two-band $k\cdot p$ Hamiltonian is diagonal. Therefore, the periodic part of the Bloch function at the $k$ points near the $\pm X$ points can be approximated as $v_{\pm X+k}(r)\approx v_{\pm X}(r)$, where $k$ is along the (001)-direction and measured from the $\pm X$ points. Under this approximation, the wave functions for the valley positions of the $\Delta_1$ and $\Delta_{2^{\prime}}$ bands in the reduced Brillouin zone can be written as:
\begin{equation}
	\begin{aligned}
		\psi_{1,\pm k_0}(\boldsymbol{r})&=u_{1,\pm k_0}(\boldsymbol{r})e^{\pm ik_0 \boldsymbol{r}}=\phi_{\pm k_0}(\boldsymbol{r})=v_{\pm k_0}(\boldsymbol{r})e^{\pm ik_0 \boldsymbol{r}}=v_{\pm (X-k_1)}(\boldsymbol{r})e^{\pm ik_0 \boldsymbol{r}} \approx  v_{\pm X}(\boldsymbol{r})e^{\pm ik_0 \boldsymbol{r}} \\
		\psi_{2,\pm k_0}(\boldsymbol{r})&=u_{2,\pm k_0}(\boldsymbol{r})e^{\pm ik_0 \boldsymbol{r}}=\phi_{\pm (k_0-\boldsymbol{G}_1)}(\boldsymbol{r})=v_{\pm (k_0-\boldsymbol{G}_1)}(\boldsymbol{r})e^{\pm i (k_0-\boldsymbol{G}_1) \boldsymbol{r}}=v_{\mp (X+k_1)}(\boldsymbol{r})e^{\pm i (k_0-\boldsymbol{G}_1) \boldsymbol{r}} \\
		& \approx  v_{\mp X}(\boldsymbol{r})e^{\pm i (k_0-\boldsymbol{G}_1) \boldsymbol{r}}.
	\end{aligned}
\end{equation}
Therefore, we can establish the connection between the periodic parts of the Bloch functions for the valley positions of the $\Delta_1$ and $\Delta_{2^{\prime}}$ bands in the reduced Brillouin zone, 
\begin{equation}
	\begin{aligned}
		u_{2,\pm k_0}(\boldsymbol{r})  \approx  v_{\mp X}(\boldsymbol{r})e^{\mp i \boldsymbol{G}_1 \boldsymbol{r}} \approx u_{1,\mp k_0}(\boldsymbol{r}) e^{\mp i \boldsymbol{G}_1 \boldsymbol{r}}.
	\end{aligned}
\end{equation}

\section{Derivation of valley coupling}
Thus, the expression for $|U_{V}|$ then reads
\begin{equation}
	\begin{aligned}
		U_{V} =&\langle \psi_{+}|H|\psi_{-}\rangle=\int_{-\infty}^{+\infty}  \Psi^*(z)u^{\prime *}_{+k_0}(\boldsymbol{r}) e^{-ik_0 z}\left[U(z)-\frac{eF}{\varepsilon}z \right] \Psi(z) u^{\prime }_{-k_0}(\boldsymbol{r}) e^{-ik_0 z} d \boldsymbol{r}\\
		=& \int_{-\infty}^{+\infty}  |\Psi(z)|^2 e^{-i2k_0 z}  \left[\cos{\left(\frac{\phi}{2}\right)} u^*_{1, + k_0} (\boldsymbol{r})-\text{sgn}(\varepsilon_{x y}) \sin{\left(\frac{\phi}{2}\right)}  u^*_{2, + k_0}(\boldsymbol{r}) \right] \cdot \left[U(z)-\frac{eF}{\varepsilon}z \right] \cdot \\
		& \left[\cos{\left(\frac{\phi}{2}\right)} u_{1, - k_0} (\boldsymbol{r})-\text{sgn}(\varepsilon_{x y}) \sin{\left(\frac{\phi}{2}\right)}  u_{2, - k_0}(\boldsymbol{r}) \right] d \boldsymbol{r}\\
		=& \int_{-\infty}^{+\infty}  |\Psi(z)|^2 e^{-i2k_0 z}  \left[ \cos^2 \left(\frac{\phi}{2}\right) u^*_{1, + k_0} (\boldsymbol{r})  u_{1, - k_0} (\boldsymbol{r})  -\text{sgn}(\varepsilon_{x y}) \sin \left(\phi\right) u^*_{1, + k_0} (\boldsymbol{r})   u_{2, - k_0}(\boldsymbol{r})   \right.\\
		& \left.    +\sin^2 \left(\frac{\phi}{2}\right)  u^*_{2, + k_0} (\boldsymbol{r})  u_{2, - k_0} (\boldsymbol{r})    \right] \cdot \left[U(z)-\frac{eF}{\varepsilon}z \right] d \boldsymbol{r}\\
	\end{aligned}
\end{equation}

We neglect higher-order strain terms proportional to $ \sin^2 \left(\frac{\phi}{2}\right)$ and substitute the plane wave expansion, yielding
\begin{equation}
	\begin{aligned}
		U_{V} \approx& \int_{-\infty}^{+\infty}  |\Psi(z)|^2 e^{-i2k_0 z}  \left[ \cos^2 \left(\frac{\phi}{2}\right) u^*_{1, + k_0} (\boldsymbol{r})  u_{1, - k_0} (\boldsymbol{r})  -\text{sgn}(\varepsilon_{x y}) \sin \left(\phi\right) u^*_{1, + k_0} (\boldsymbol{r})   u_{2, - k_0}(\boldsymbol{r})   \right] \cdot \left[U(z)-\frac{eF}{\varepsilon}z \right] d \boldsymbol{r}\\
		\approx&  \int_{-\infty}^{+\infty}  |\Psi(z)|^2 e^{-i2k_0 z}  \left[U(z)-\frac{eF}{\varepsilon}z \right] \cdot \left[ \cos^2 \left(\frac{\phi}{2}\right) \sum_{\boldsymbol{G}}   c^*_+(\boldsymbol{G})e^{-i\boldsymbol{G}\cdot \boldsymbol{r}}  \sum_{\boldsymbol{G}^\prime}   c_-(\boldsymbol{G}^\prime)e^{i\boldsymbol{G}^\prime\cdot \boldsymbol{r}} -\text{sgn}(\varepsilon_{x y})   \sin \left(\phi\right)  \cdot  \right.   \\
		&\left.    \sum_{\boldsymbol{G}}   c^*_+(\boldsymbol{G})e^{-i\boldsymbol{G}\cdot \boldsymbol{r}}    \sum_{\boldsymbol{G}^{\prime \prime}}   c_+(\boldsymbol{G}^{\prime \prime})e^{i(\boldsymbol{G}^{\prime \prime}+\boldsymbol{G}_1)\cdot \boldsymbol{r}} \right] d \boldsymbol{r}\\
		=& \sum_{\boldsymbol{G},\boldsymbol{G^\prime}}\delta(G_x^\prime-G_x)\delta(G_y^\prime-G_y)  \cos^2(\frac{\phi}{2})  c_+^*(\boldsymbol{G})c_-(\boldsymbol{G^\prime}) I_1(G_z,G_z^\prime) - \sum_{\boldsymbol{G},\boldsymbol{G^{\prime \prime}}}\delta(G_x^{\prime \prime}-G_x)\delta(G_y^{\prime \prime}-G_y) \text{sgn}(\varepsilon_{x y})   \sin \left(\phi\right) \cdot \\
		&  c^*_+(\boldsymbol{G}) c_+(\boldsymbol{G}^{\prime \prime}) I_2(G_z,G_z^{\prime \prime}), \\
	\end{aligned}
	\label{eq-s4}
\end{equation}
where the $I_n$ terms stand for the integral
\begin{equation}
	\begin{aligned}
		I_1(G_z,G_z^\prime)&=\int^{+\infty}_{-\infty}|\Psi(z)|^2 e^{i(G_z^\prime-G_z-2k_0) z}\left[U(z)-\frac{eF}{\epsilon}z \right]dz  \\
		I_2(G_z,G_z^{\prime \prime})&=\int^{+\infty}_{-\infty}|\Psi(z)|^2 e^{i(G_z^{\prime \prime}-G_z+2k_1) z}\left[U(z)-\frac{eF}{\epsilon}z \right]dz.  \\
	\end{aligned}
	\label{eq-s5}
\end{equation}
We approximate \( U(z) \) as a finite potential well with a barrier height \( U_0 = 150 \, \text{meV} \) and boundaries at \( \pm L/2 \), i.e.,
\begin{equation}
	U(z) = \begin{cases} 
		U_0, & |z| > \frac{L}{2}, \\
		0, & |z| \leq \frac{L}{2}. 
	\end{cases}
\end{equation}
An electric field \( Fz / \epsilon \) along the \( z \)-direction confines the electron near the upper interface. Thus, we can neglect the contribution of the lower interface and  integrate by parts the term proportional to $U(z)$ in Eq.~\ref{eq-s5},
\begin{equation}
	\begin{aligned}
		I_1(G_z,G_z^\prime)&=\frac{i}{Q_1}U_0\left|\Psi(z_i) \right|^2 e^{iQ_1z_i} +\int_{L/2}^{+\infty}\frac{1}{Q_1}U_0 \frac{d\left|\Psi(z)\right|^2}{dz} e^{iQ_1z}dz-\int_{-\infty}^{+\infty}\left|\Psi(z)\right|^2e^{iQ_1z}\frac{eF}{\epsilon} zdz\\
		I_2(G_z,G_z^{\prime \prime})&=\frac{i}{Q_2}U_0\left|\Psi(z_i) \right|^2 e^{iQ_2z_i} +\int_{L/2}^{+\infty}\frac{1}{Q_2}U_0 \frac{d\left|\Psi(z)\right|^2}{dz} e^{iQ_2z}dz-\int_{-\infty}^{+\infty}\left|\Psi(z)\right|^2e^{iQ_2z}\frac{eF}{\epsilon} zdz  , \\
	\end{aligned}
\end{equation}
where $Q_1=G_z^\prime-G_z-2k_0$, and $Q_2=G_z^{\prime\prime}-G_z+2k_1$. The last two terms, originating from the evanescent tail of the electronic envelope function into the barrier material and intervalley scattering induced directly by the electric field, are found to be negligibly small. Therefore, we can consider only the contribution from the electronic density at the interface in the first term. Moreover, only terms with $G_z=G_z^\prime$ ($G_z=G_z^{\prime\prime}$) need to be considered, as those with $G_z\neq G_z^\prime$ ($G_z\neq G_z^{\prime\prime}$) lead to rapid oscillations that average to zero in the integrand. Thus, Eq.~\ref{eq-s4} can be further simplified to
\begin{equation}
	\begin{aligned}
		U_{V} 		=& -\sum_{\boldsymbol{G}} c_+^*(\boldsymbol{G})c_-(\boldsymbol{G})  \frac{i\cos^2(\frac{\phi}{2}) }{2k_0}U_0\left|\Psi(z_i) \right|^2 e^{-2ik_0 z_i}- \sum_{\boldsymbol{G}}  c^*_+(\boldsymbol{G}) c_+(\boldsymbol{G})  \frac{i~\text{sgn}(\varepsilon_{x y})   \sin \left(\phi\right) }{2k_1}U_0\left|\Psi(z_i) \right|^2 e^{2ik_1 z_i} \\
		=& \cos^2\left(\frac{\phi}{2}\right) \frac{ \Sigma_1 U_0}{i2k_0}   \left|\Psi(z_i) \right|^2 e^{-2ik_0 z_i} + \text{sgn}(\varepsilon_{x y}) \sin \left(\phi\right) \frac{ \Sigma_2 U_0 }{i2k_1}   \left|\Psi(z_i) \right|^2 e^{2ik_1 z_i }
	\end{aligned}
\end{equation}
where $\Sigma_1=\sum_G c_+^*(\boldsymbol{G})c_-(\boldsymbol{G})=-0.2607$, and $\Sigma_2=\sum_{\boldsymbol{G}} c_+^*(\boldsymbol{G})c_+(\boldsymbol{G})=0.9954$, as based on Ref.~\cite{PhysRevB.84.155320}.

\section{Wave function}
\begin{figure}[t]
	\centering
	\includegraphics[width=0.8\textwidth]{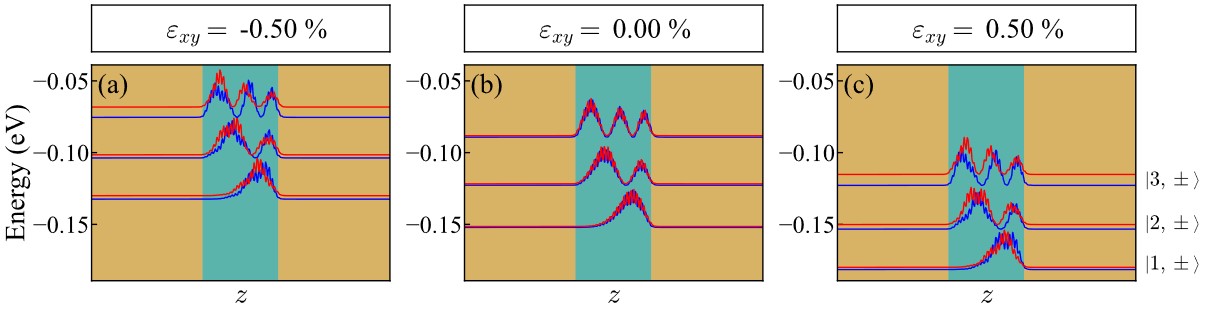}
	\caption{ Wave functions of six lowest electronic bound states in a Si(53 ML)/Ge QW subjected to an electric field of $10~\mathrm{MV} / \mathrm{m}$ along the [001] direction. Panels (a-c) depict Si/Ge QW structures at different shear strains. The red and blue lines at specific energy levels represent valley pairs of different orbital states $|n, \pm\rangle$. The zero energy reference is set to the CBM of bulk Si.
		\label{figS1}
	}
\end{figure}

Shear strain profoundly influences the electronic bound states within Si/Ge QWs. Depicted in Fig.~\ref{figS1}(a) and (c) are the electronic bound states in Si/Ge QWs subjected to compressive and tensile shear strains, respectively. Fig.~\ref{figS1}(b) illustrates the electronic bound states in Si/Ge QWs without shear strain. As the shear strain transitions from negative to positive (i.e., from compressive to tensile shear strain), the confinement energy of the Si/Ge QW intensifies, resulting in a reduction of the ground state energy and an augmentation in the number of electronic bound states. Additionally, higher orbital states exhibit larger $E_\text{VS}$.

\begin{table}[t]
	\centering
	\caption{Values of $\theta$ for QWs containing $4n$, $4n+1$, $4n+2$, and $4n+3$ Si MLs under positive and negative electric fields, considering the ideal interface positions located half an atomic layer on either side of the QW [Fig.~\ref{figS2}(a)].}
	\begin{tabular}{lrrc}
		\hline
		\hline
		\bf{Si MLs} & \bf{Upper} & \bf{Lower}  & \bf{Field-reversal dependence} \\ \hline
		4n    & $\pi$/2    & $\pi$/2    & \ding{55}  \\
		4n+1  & $\pi$/2    & -$\pi$/2   & \ding{51}\\
		4n+2  & -$\pi$/2   & -$\pi$/2  & \ding{55} \\
		4n+3  & -$\pi$/2   & $\pi$/2    & \ding{51}\\ \hline\hline
	\end{tabular}
	\label{tabS1}
\end{table}

\begin{figure}[htp]
	\centering
	\includegraphics[width=1.0\textwidth]{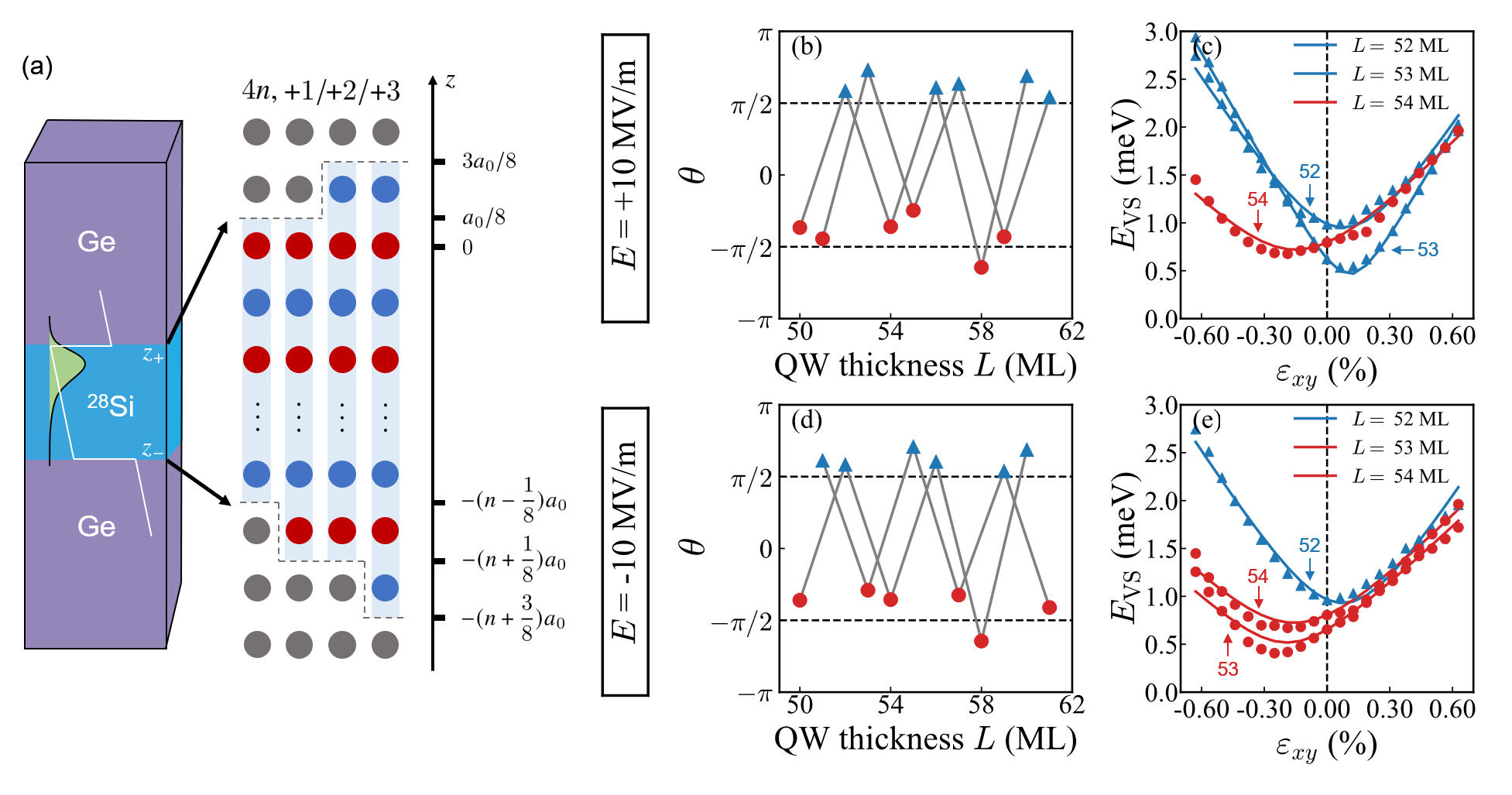}
	\caption{ (a) Schematic illustration of a Si/Ge QW with enlarged views of the upper and lower interfaces for $4n$, $4n+1$, $4n+2$ and $4n+3$ Si MLs. The figure shows a positive electric field pushing electrons toward the upper interface, and the reverse electric field would push electrons toward the lower interface. The coordinates of the ideal interface are shown on the right, with the uppermost Si atomic layer of the (4n)-ML QW taken as the origin. (b) Variation of the phase difference between inter-BZ $\left(V_{+-}^X\right)$ and intra-BZ $\left(V_{+-}^{\Gamma}\right)$ valley couplings in an ideal Si/Ge QW as a function of well thickness. The data is obtained by fitting Eq.~\ref{eq.vs} to results from atomistic SEPM simulations. (c) Variation of $E_{\mathrm{VS}}$ with shear strain in 52 ML, 53 ML, and 54 ML thick ideal Si/Ge QWs. The discrete data points represent results from atomistic SEPM calculations, while the curves represent the fitting results. In (b) and (c), a uniform electric field of $10~\mathrm{MV} / \mathrm{m}$ was applied, pushing electrons toward the upper interface of the QW. (d) and (e) present the corresponding results under a reversed electric field of $-10~\mathrm{MV} / \mathrm{m}$, which shifts the electron wavefunction toward the lower interface of the QW.
		\label{figS2}
	}
\end{figure}

\section{Validation of the model through ideal quantum well simulations}
To validate the proposed valley splitting model, we performed atomistic simulations for ideal Si/Ge QWs with varying thicknesses under positive and negative electric fields. Our analysis focuses on the influence of QW thickness and electric field direction---which determine the interface position ($z_i$)---on the relative phase angle $\theta$ and the trend of valley splitting under shear strain.

We begin by examining the model predictions. As defined in Eq.~\ref{eq.vs}, the relative phase angle $\theta$ depends sensitively on the interface position $z_i$. However, due to the intrinsic ambiguity of the atomic-scale interface location, $z_i$ may vary within the range of a single monolayer. Previous studies have suggested that the effective scattering interface in Si/Ge QWs typically lies slightly offset from the midpoint between two atomic planes, as illustrated in Fig.~\ref{figS2}(a), and must be inferred from atomistic calculations~\cite{PhysRevB.75.115318}. Table~\ref{tabS1} summarizes the values of $\theta$ for QWs of thickness of $4n$, $4n+1$, $4n+2$ and $4n+3$ MLs, assuming the interface lies exactly midway between two atomic layers [with coordinates indicated in Fig.~\ref{figS2}(a)]. The results show that $\theta$ oscillates periodically between 
$\pi/2$ and $-\pi/2$ with QW thickness, and its sign changes under electric field reversal only for QWs with $4n+1$ or 
$4n+3$ MLs. Under this simplified assumption for the interface location, the intra- and inter-BZ valley coupling terms are decoupled due to $\cos (\pm \pi/2)=0$. However, in real structures, deviations from the ideal interface position lead to a slight shift in $\theta$ away from $\pm \pi/2$. Since $E_{\mathrm{VS}}$ reaches a minimum when $V_{+-}^X=-V_{+-}^\Gamma \cos \theta$, any deviation from the ideal relative phase angle introduces asymmetry in the strain dependence of valley splitting. We demonstrate below that the atomistic simulation results are in full agreement with these model predictions, confirming its validity.

First, we analyze the oscillatory behavior of the relative phase angle as a function of interface position under a positive electric field. As shown in Fig.~\ref{figS2}(b), the fitted values of $\theta$ exhibit a clear dependence on QW thickness, forming two branches corresponding to the two sublattices of the diamond crystal structure (offset by one ML, leading to a $\pi$ shift in $\theta$).  Specifically, QWs with $4n$ and $4n+1$ MLs have $\theta \approx \pi/2$, while those with $4n+2$ and $4n+3$ MLs exhibit $\theta \approx -\pi/2$, consistent with Table~\ref{tabS1}. Since the sign of $\cos \theta$ determines whether $E_{\mathrm{VS}}$ reaches its minimum under tensile or compressive strain, small deviations in $\theta$ from $\pm \pi/2$ explain the anomalous behavior of $E_{\mathrm{VS}}$ under weak strain in Fig.~\ref{figS2}(c). For example, in an 52-ML QW, $\cos \theta<0$ leads to an initial decrease in $E_{\mathrm{VS}}$ under tensile strain before it increases, whereas in an 54-ML QW, $\cos \theta>0$ results in a similar trend under compressive strain.

Second, we examine the change of nonmonotonic response under electric field reversal. Due to thickness-dependent variation in the relative positions of the top and bottom interfaces, the associated phase factor also differs. To assess this effect, we reversed the electric field and compared the resulting changes in $\theta$ and $E_{\mathrm{VS}}$. As illustrated in Fig.~\ref{figS2}(d), $\theta$ remains nearly unchanged for QWs with $4n$ and $4n+2$ MLs under field reversal, while for $4n+1$ and $4n+3$ MLs, $\theta$ flips sign. This behavior is in line with the model prediction: only odd-ML QWs (with $D_{2d}$ symmetry) are sensitive to electric field polarity, whereas even-ML QWs (with $D_{2h}$ symmetry) are not. Consequently, as shown in Fig.~\ref{figS2}(e), the $E_{\mathrm{VS}}$-strain relation for the 53-ML QW changes significantly under field reversal, while 52-ML and 54-ML QWs show minimal variation.

These findings indicate that minor deviations of the effective scattering interface from the ideal location result in $\theta$ shifting away from $\pm \pi/2$, introducing complex interference between intra- and inter-BZ valley couplings. This not only validates the proposed model’s capability in capturing shear strain-induced valley splitting, but also highlights the non-monotonic strain dependence of valley splitting in ideal QWs, driven by atomic-scale uncertainties in interface positioning. The results also underscore the importance of atomistic simulations in elucidating valley physics in Si/Ge QWs.

\section{Effect of Increased Step Density on Valley Coupling}
\begin{figure}[t]
	\centering
	\includegraphics[width=1.0\textwidth]{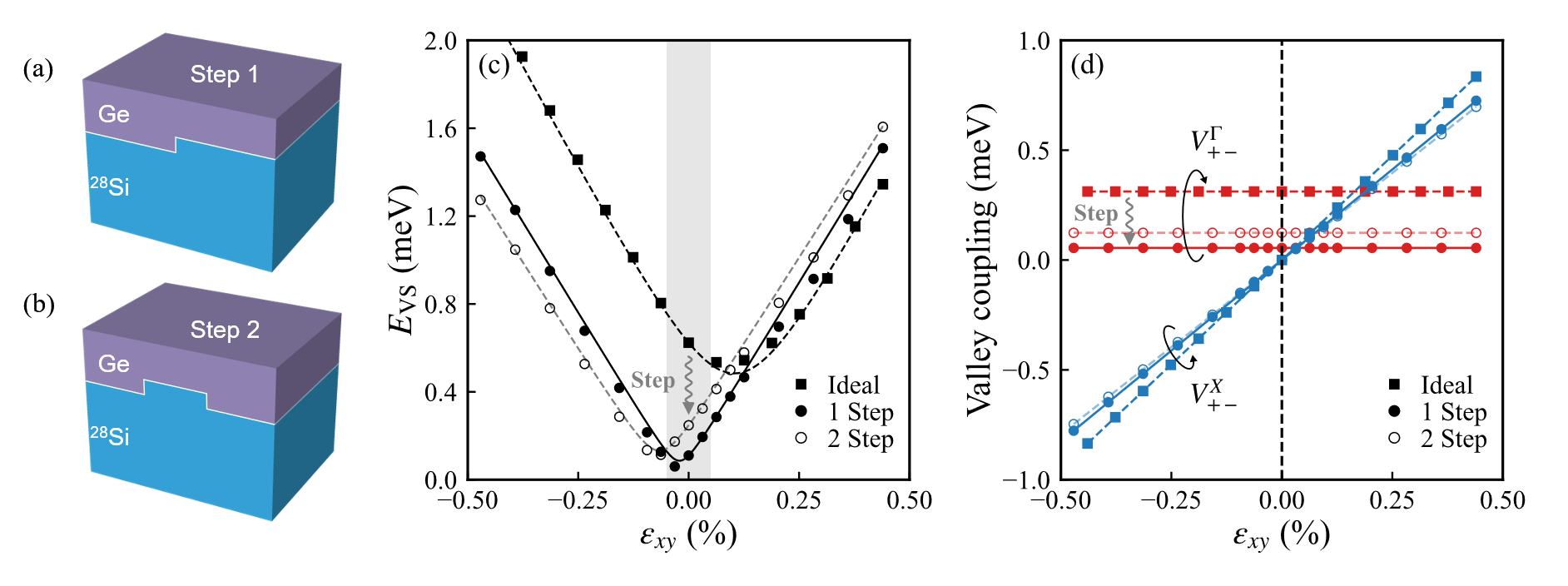}
	\caption{(a) Single-step model in main text. Schematic representation of the top interface of the Si/Ge QW, featuring a single atomic step with a height of 1 ML. The step edge is oriented along the [010] direction and centered along the [100] direction on the (001) interface plane. (b) Schematic representation of a modified interface configuration with doubled step density. Two equally spaced 1-ML-high atomic steps are introduced along the [100] direction, while maintaining the same overall geometry and orientation as in (a). (c) Valley splitting energy in (001)-oriented $53~\mathrm{ML}$-$\mathrm{Si}/\mathrm{Ge}$ QWs calculated using the atomistic SEPM. Black solid squares represent results for $\mathrm{Si} / \mathrm{Ge}$ QWs with atomically flat interfaces, while black circles depict the case with interface steps. The solid/dashed lines represent fits based on Eq.~\ref{eq.vs}. (d) Dependence of inter-BZ valley coupling $V_{+-}^X$ (blue) and intra-BZ valley couplings $V_{+-}^{\Gamma}$ (red) on shear strain in a 53 ML-thick Si QW. Solid squares represent the scenario of an ideal Si QW, and circles correspond to Si QWs with the interface step. An uniform electric field of $10~\mathrm{MV} / \mathrm{m}$ is applied during the atomistic SEPM simulation.
		\label{figS3}
	}
\end{figure}

To further validate the generality of our findings, we also analyze a second configuration [Fig.~\ref{figS3}(b)] in which the step face density is doubled relative to the first configuration in the main text [Fig.~\ref{figS3}(a)]. This configuration preserves the overall geometry but introduces two equally spaced 1-ML steps along the [100] direction. The doubled step density divides the QW interface into two regions with a 2:1 width ratio. As a result, the phase-related reduction factors for the intra- and inter-BZ valley coupling components become $\eta^\Gamma=\left|\left(2+e^{-ik_0 a_0/2}\right)/3\right|=39.94\%$, and  $ \eta^X=\left|\left(2+e^{ik_1 a_0/2}\right)/3\right|=97.54\%$. Compared to the first configuration, the inter-BZ valley coupling remains nearly unchanged, while the intra-BZ component exhibits a slightly weaker suppression. This observation aligns with the SEPM results shown in Figs.~\ref{figS3}(c) and (d), where three key features match expectations: the fitted value of $\eta^\Gamma$ increases from 0.055 meV to 0.124 meV, $\eta^X$ exhibits a similar trend as in the single-step configuration, and the shear strain at which the valley splitting reaches its minimum shifts from $-0.03\%$ to $-0.06\%$. These results further reinforce the robustness of the underlying mechanism governing valley coupling behavior in the presence of interfacial disorder.

\end{document}